\documentclass[conference]{IEEEtran}
\IEEEoverridecommandlockouts
\usepackage{cite}
\usepackage{amsmath,amssymb,amsfonts}
\usepackage{algorithmic}
\usepackage{lipsum}
\usepackage{graphicx}
\usepackage{hyperref}
\usepackage{graphicx}
\usepackage{svg}
\usepackage{textcomp}\usepackage{xcolor}
\def\BibTeX{{\rm B\kern-.05em{\sc i\kern-.025em b}\kern-.08em
    T\kern-.1667em\lower.7ex\hbox{E}\kern-.125emX}}
\begin{document}
\newcommand\blfootnote[1]{%
  \begingroup
  \renewcommand\thefootnote{}\footnote{#1}%
  \addtocounter{footnote}{-1}%
  \endgroup
}
% Department of Computer Science, Mathematics, Physics and Statistics
\newcommand{\KTDis}{Kendall-Tau distance}
\title{Metric Search for Rank List Compatibility Matching with Applications
}

\author{

\IEEEauthorblockN{{\bf Wenqi Guo}*\thanks{{\em *Work was initiated while the author was an undergraduate research student at the University of Missouri-Columbia}.}}
\IEEEauthorblockA{{Department of CMPS} \\
{University of British Columbia}\\
Kelowna, BC, Canada\\
ORCID: 0000-0001-6125-0255
}

\and
\IEEEauthorblockN{\bf Jeffrey Uhlmann}
\IEEEauthorblockA{{Department of EECS} \\
{University of Missouri}\\
Columbia, MO, USA \\
ORCID: 0000-0001-9163-521X
}
}

\maketitle
\begin{abstract}
% In this paper, we consider the use of metric search with
% the Kendall-Tau metric to identify similarity among
% lists of ranked tokens efficiently. Specifically, we propose and examine
% the use of rank similarity matching as the core algorithm for
% compatibility assessment, e.g., for dating apps. We then provide
% empirical results showing that metric search can be used to
% efficiently satisfy rank list similarity queries on problems of
% practical size. 
As online dating has become more popular in the past few years, an efficient and effective algorithm to match users is needed. In this project, we proposed a new dating matching algorithm that uses Kendall-Tau distance to measure the similarity between users based on their ranking for items in a list. (e.g., their favourite sports, music, etc.) To increase the performance of the search process, we applied a tree-based searching structure, Cascading Metric Tree (CMT), on this metric. The tree is built on ranked lists from all the users; when a query target and a radius are provided, our algorithm can return users within the radius of the target. We tested the scaling of this searching method on a synthetic dataset by varying list length, population size, and query radius. We observed that the algorithm is able to query the best matching people for the user in a practical time, given reasonable parameters. We also provided potential future improvements that can be made to this algorithm based on the limitations. Finally, we offered more use cases of this search structure on Kendall-Tau distance and new insight into real-world applications of distance search structures. 
\end{abstract}

\section{Introduction}
Measuring the compatibility between two people could be a complex task. 
Existing dating applications Tinder \cite{Tinder} does not collect much information from the users. Instead, it returns a set of profiles within the geographic radius to the users that satisfy the user's query conditions (age range and gender). Users can choose either \textit{Yes} or \textit{No} to this profile. \cite{TinderResearch} If there is a mutual like, there is a ``match'' between the two users, and they can start chatting. Some other dating platforms, such as Hinge \cite{Hinge}, use an algorithm based on each user's interaction and match users with people with similar preferences. \cite{HingeAl} \cite{HingeResearch} However, both methods are appearance-focused and do not rely much on users' profiles, which might not lead to a successful relationship. 

The questionnaire is another common dating matching method that has been used for a long time before Internet dating existed. One of the questionnaires is to collect people's interests and their expectations for their partners, and a dating agency will match people's profiles and arrange for dates. 

In this paper, we proposed a similar algorithm that allows users to provide a rank list of their interests, such as favourite movies or sports. The algorithm will then measure the compatibility between individuals based on their rank list and match people with similar lists. 
%Although because the algorithm of Tinder is not open source 

\begin{figure}[]
Measuring compatibility is important in dating applications
\centering
\includegraphics[width=0.3\textwidth]{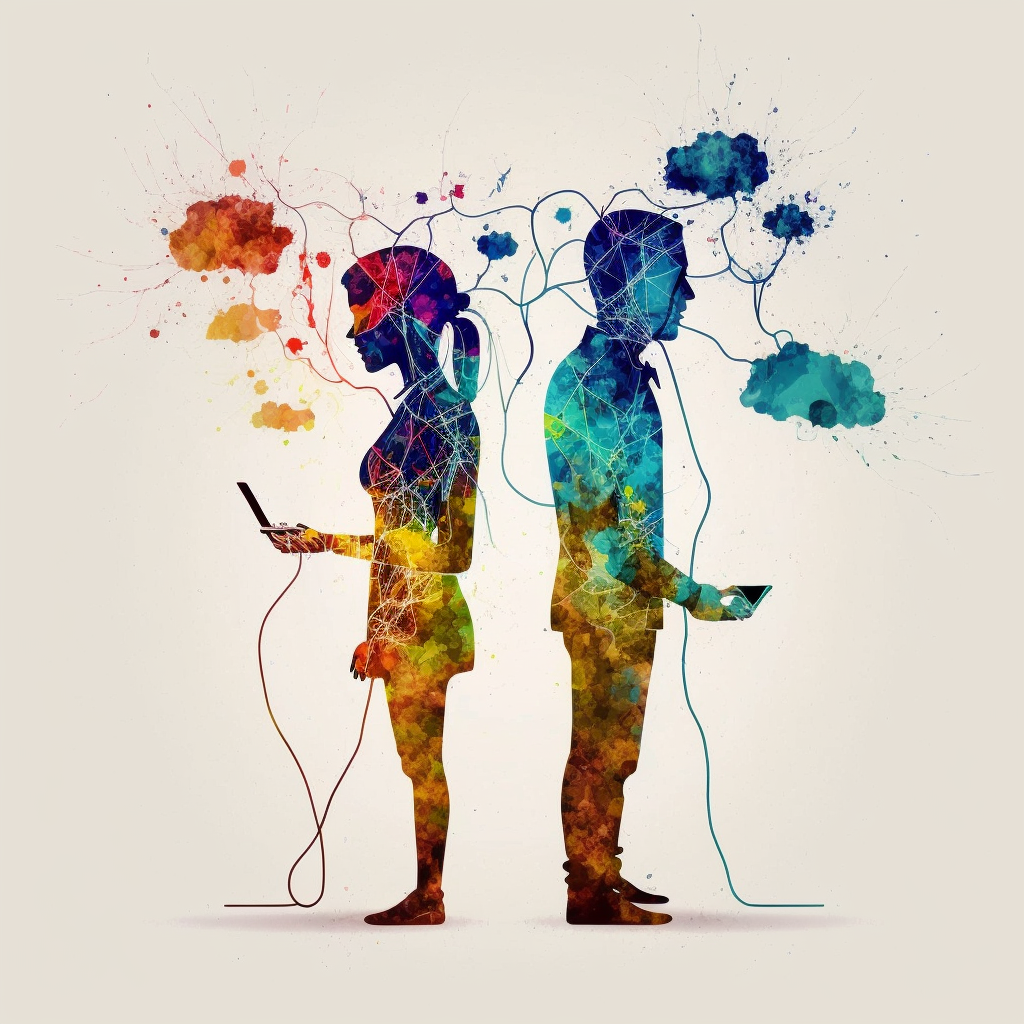}
\end{figure}

\section{The Kendall-Tau Metric and Compatibility Measures}
Kendall Tau Distance measures the number of inversions of pairs between two ordered lists, or it could be described as the number of adjacent swaps that need to make from one list to another. It is a metric modified from Kendall-Tau Correlation. \cite{KTD}\cite{KTS} 

Example: 
Suppose we have two ranked lists, $\tau_1 = [A, B, C, D]$ and $\tau_2 = [C, D, A, B]$. We define $A>B$ as \textit{A is prior than B in the list}, and all the pairs and their orders in each list can be represented as in Table 1. The number of pair order disagreements is 4. Thus, the Kendall Tau Distance between $\tau_1$ and $\tau_2$. We could normalize this distance to make its range between 0 and 1 for ease to query later by dividing the distance by the number of all pairs (Which can be computed by $(n-1)n/2$). \cite[p. 4]{KTS} Thus, the normorlized Kendall Tau Distance is $\frac{4}{4(4-1)/2}=\frac{2}{3}$.
In our dating app use cases, we let users rank items based on their interests, for example, their favourite sports or movies. With the rank lists between people, the \KTDis{} could be used to measure the similarity between the two people. The more similar two individuals are, we say they are more ``compatibl'', which means they are more likely to have a successful relationship. Therefore, we can use the \KTDis{} as the compatibility measures between two users: the larger the distance, the less compatible the two individuals are. 

\section{Metric Search Structures}
Due to the fact that it is a metric, Kendall-Tau distance has the following properties:\cite[p. 2]{KTS} \cite[p. 2]{CMT}
$$
\forall i,j,k \in S \textbf{ where $S$ is the object set}
$$
$$
\left\{
\begin{array}{cc}
d(i, j) \geq 0 &  \text{(Non-negativity)}\\
d(i, j)=0 \Leftrightarrow i = j & \text{(Identity of Indiscernibles)}\\
d(i, j) =  d(j, i) & \text{ (Symmetry)}\\
d(i, j) + d(j, k) \leq d(i, k)  & \text{ (Triangle Inequality)}
\end{array}
\right.$$

Since we have a distance function to measure the compatibility between two people, a straightforward way to select dating candidates is to iterate through each user in the database and calculate the distance between such user and the querying user. Then, we could sort the result based on the distance and select users that are closer to the querying user as dating candidates for the querying user. However, this algorithm has a time complexity of $\mathcal{O}(N)$, where N is the total number of users in the database. Therefore, considering the massive number of users, this approach cannot scale efficiently with $N$. 

To increase the performance of the search algorithm, a metric tree can be used. When constructing a metic tree, given a set of objects, a random object was selected in such a set as the root object. Then the distances between all other objects and the root object are calculated. These objects are separated into two sets $Inner$ and $Outer$, and $\forall i \in Inner, \forall j \in Outer;\:d(root, i)<d(root, j)$. Ideally, the cardinalities of these two sets should be the same or at least similar. The minimum and maximum distance between the root and all objects ($r_{min} = \min_{i\in Objects} d(i, root)$ and $r_{max}=\max_{i\in Objects} d(i, root)$,\; respectively), and the maximum distance between the root and the inner set ($r_{inner} = \max_{i\in Inner} d(i, root)$) and the minimum distance between the root and the outer set ($r_{outer} = \min_{i\in Outer} d(i, root)$)are stored at the root node. Then, we recursively construct the left and right subtree using the $Inner$ and $Outer$ sets as object sets. \cite{CMT} \cite{OldMT} For convenience, we define agree that the $Inner$ is passed to the left subtree and $Outer$ is passed to the right subtree in this article.

In the query process, an object $q$ and a radius $r_q$ are given. We define the query ball as a sphere with a center $q$ and a radius $r_q$ containing the target query space. For each node, there are four possible conditions: \cite{CMT}
\begin{itemize}
    \item \textbf{If $d(root, q)<r_{q}$ (Cond. 0)} 
        \subitem This implies that the root object is within the query ball; thus, it should be added to the answer set.
    \item \textbf{If $d(root, q)+r_q<r_{min}$ (Cond. 1)} 
        \subitem This implies that the query ball is inside the minimum boundary of all objects in the tree; thus, no objects will fall into the query ball, and we can prune this branch. (Cond. 2-4 will not be checked)
    \item \textbf{If $d(root, q)-r_q>r_{max}$ (Cond. 2)} 
        \subitem This implies that the query ball is outside the maximum boundary of the tree, and we can prune this branch. (Cond. 3-4 will not be checked)
    \item \textbf{If $d(root, q)-r_q<r_{inner}$ (Cond. 3)} 
        \subitem This implies that the query ball intersected with the left subtree; thus, recursively search the left subtree if it is not empty. 
    \item \textbf{If $d(root, q)-r_q>r_{outer}$ (Cond. 4)} 
        \subitem This implies that the query ball intersected with the left subtree; thus, recursively search the left subtree if it is not empty. 
\end{itemize}
(Note that Condition 3 and 4 are not mutually exclusive because the query ball can intersect with both the left and right subtree.) 

The time complexity of this algorithm at search time is $\mathcal{O}(log N)$, meaning it performs about $\mathcal{O}(log N)$ distance calculations for each query. \cite{CMT} However, as $r\rightarrow 1$ (Since our normalized \KTDis{} is between 0 and 1.), the number of calls to the distance function approaches N. To decrease the  number of distance calls
%, other pruning tests can be included. 
, \cite{CMT} proposed a method that stores cascading information at each node to provide additional pruning tests.  
\begin{figure*}[t]
    \centering
    \includegraphics[width=0.7\textwidth]{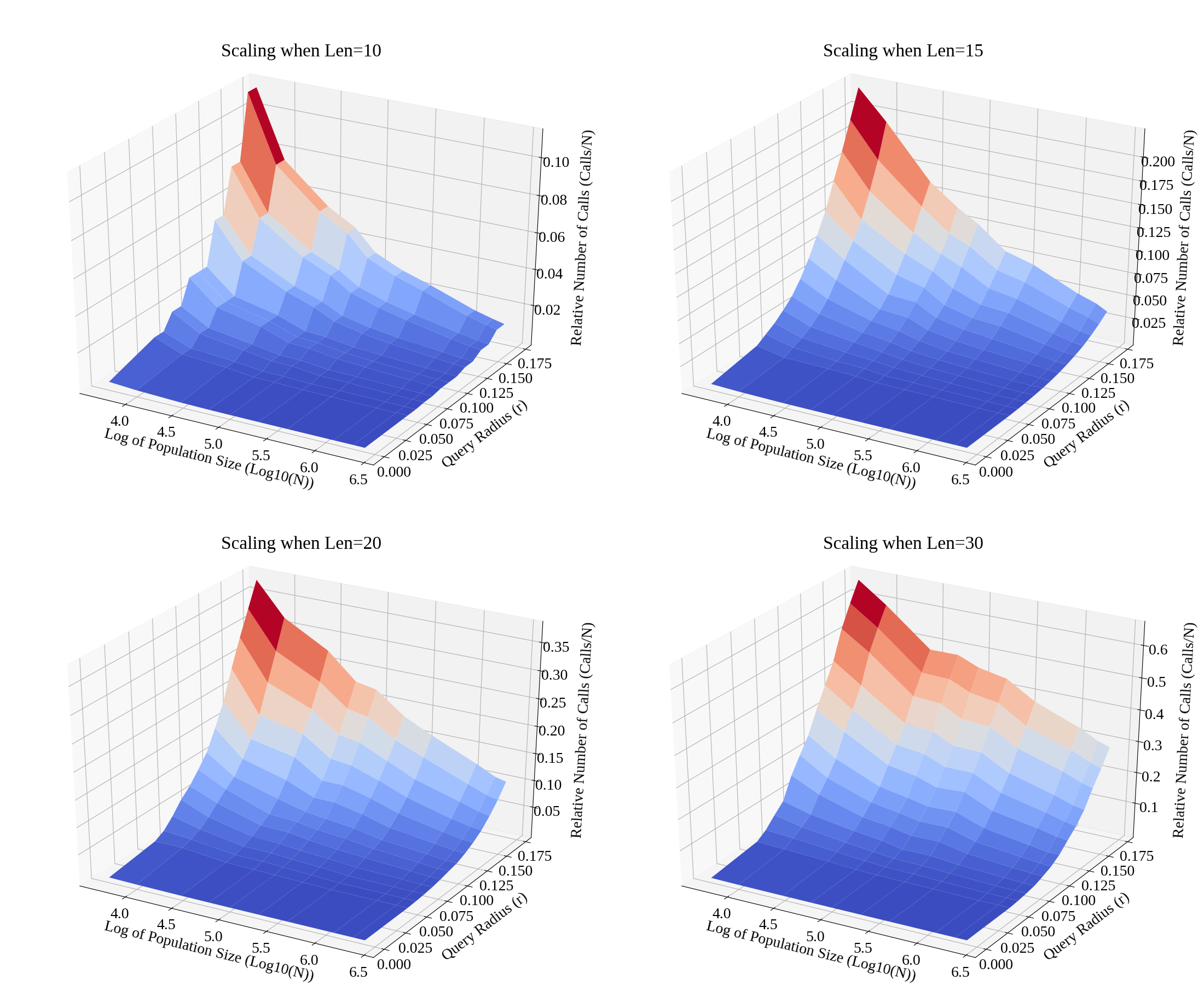}
    \caption{The scaling of the search structure based on $N$ and $r$ at different $len$}
\end{figure*}
\begin{figure*}[t]
    \centering
    \includegraphics[width=\textwidth]{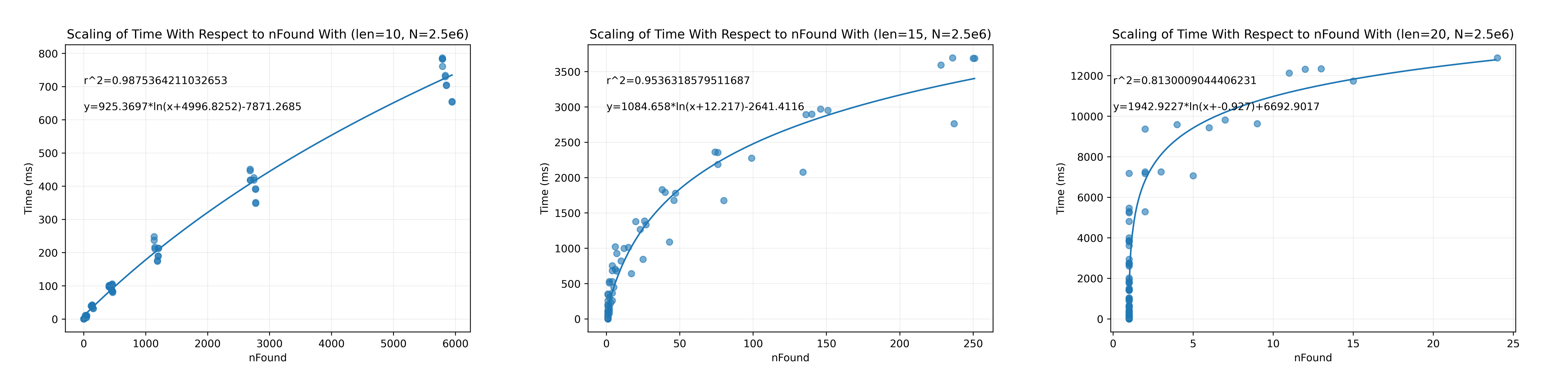}
    \caption{The scaling of the search structure based on $nFound$ at different $len$}
\end{figure*}

\begin{figure*}[t]
    \centering
    \includegraphics[width=\textwidth]{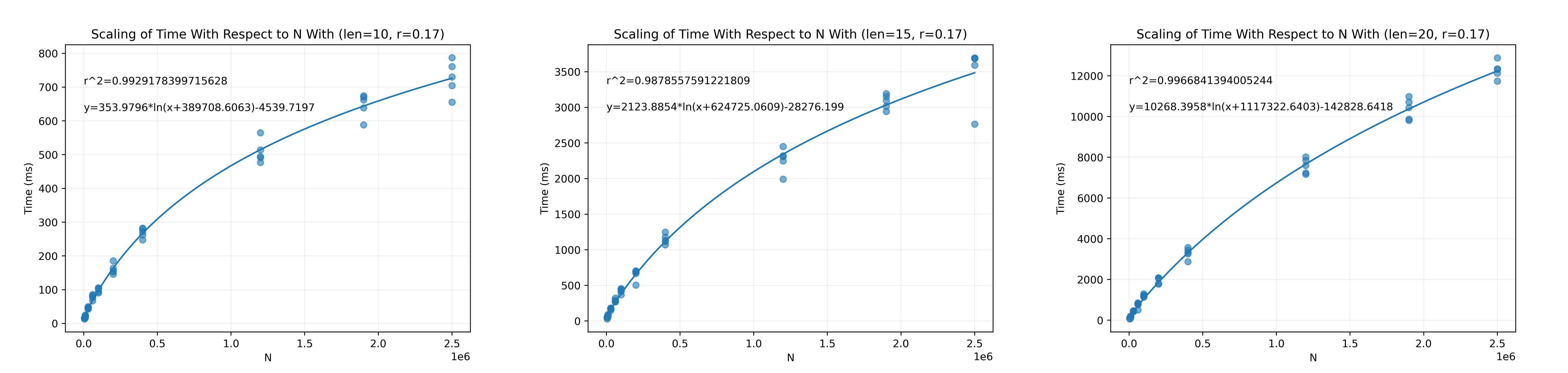}
    \caption{The scaling of the search structure based on $N$ at different $len$}
\end{figure*} 
\begin{table}
\caption{Example For Kendall Tau Distance Calculation}
\begin{center}
\begin{tabular}{|c|c|c|c|}
\hline
\textbf{Pair} & \textbf{Order In $\tau_1$}  & \textbf{Order in $\tau_2$}  & \textbf{Disagreement}
\\\hline
AB  & $A>B$  & $A>B$ & False 
\\\hline
AC   & $A>C$  & $C>A$ & True 
\\\hline
AD & $A>D$  & $D>A$ & True 
\\\hline
BC   & $B>C$  & $C>B$ & True 
\\\hline
BD & $B>C$  & $D>B$ & True 
\\\hline
CD& $C>D$  & $C>D$ & False 
\\\hline 
\end{tabular}
\label{tab1}
\end{center}
\end{table}
%simlar

\section{Methods}
\begin{table}
\caption{Testing Parameters}
\begin{center}
\begin{tabular}{|c|c|}
\hline
$r$ & $\{0.0, 0.05, 0.06 \dots, 0.16, 0.17\}$\\
\hline
$N$ & \{5e3, 1e4, 3e4, 6e4, 1e5, 2e5, 4e5,1.2e6, 1.9e6, 2.5e6\}\\
\hline
$len$ & \{10,15,20,30\} \\
\hline

\end{tabular}
\label{tab2}
\end{center}
\end{table}
We implemented the Kendall-Tau Distance using Merge-Sort in C++ based on \cite{DistanceAnswer1}, and \cite{DistanceAnswer2}. \footnote{\textit{ The implementation of the algorithm, raw test data, and data processing notebook can be found on \url{https://github.com/weathon/datinng-app-pub}}}
The implementation was 
tested by comparing it with a brute-force implementation. 

We then implemented the CMT as described in \textit{Metric Tree} section and tested the search structure for all the combinations of different list lengths ($len$), data set population size ($N$), and query radii ($r$), as shown in Table 2. For each set of parameters, $N$ lists were generated, and each list is a shuffled copy of $\{0, 1, \ldots, len-1\}$,  %WC
then five constant query objects that are in the population set were tested. For each query object, the 
number of objects found in the returned set ($nFound$), time consumed ($Time$), and number of calls to the \KTDis{} function ($Call\; Counts$) was recorded.  

The specifications of the benchmarks platform are:
\begin{itemize}
\item Intel(R) Xeon(R) CPU E5-2670 v3 @ 2.30GHz (For each test, only one core was used)
\item C++ GCC 10.2.0 on Linux without -O flag
\item Max RAM Usage: 4.498GB
\end{itemize}

\section{Results and Discussion}

Figure 1 shows the scaling of our structure with different parameters. Note that the axis for the population is in its log form, and the z-axis is the relative number of distance evaluations, i.e. the number of evaluations divided by the population size. These plots show that in all the parameters we tested, there is a significant performance advantage compared to brute force (which has a relative distance evaluation of 1). 
We observed that as $N$ increases, the relative distance evaluation decreases, which indicates that our algorithm has more advantages on larger population datasets. The relative call counts also increase superlinearly with respect to the length of the ranked list, suggesting that in real applications, multiple short lists might be a better performance choice than a long list. The call counts also increase as $r$ increase; this is as expected as queries with higher $r$ return more objects. 
% However, without knowing the number of objects returned ($nFound$), no discussion can be made on the algorithm's performance with respect to $r$. 

However, to better assess the performance of the search structure, more real-world related data ($nFound$ and $Time$) is needed. Since the graph has similar shapes across different $N$ and $len$, we only plotted the data with the fixed $N=2.5e6$ and $r=0.17$ (The worst situation we tested), as shown in Figure 2. The parameter sets with $len=30$ are excluded from the plot because we found that when $len=30$, the $nFound$ is always 1 for all the parameters we tested. This implies when $len$ is large; the objects are far away from each other even after normalization. 

From figure 2, we can again observe the search structure scaling badly with respect to $len$. Not only it took more time to search, but it also found fewer objects. However, with reasonable rank list length ($len\leq 20$), all the tests can be executed in under 4 seconds.

Figure 3 shows the scaling of the algorithm with respect to $N$. We can see that the time also increases sublinearly as $N$ increases. 
% This means when the user base grows, the query time for each additional user decreases. 
This also made it possible for users to change their rank list. When a user changes their rank list, the old data point will not be deleted since it will serve as a jump board in the tree for other users; however, it will be marked as inactive. Then, a new data point with the user's new interest will be added to the tree. Since our algorithm scales sublinear to $N$, this will not add much overhead. (The tree could be cleaned occasionally to remove these points.)

% \section{Conclusions}
% From the  
% since the time for each call is the same for each call, we can conclude that the most time is due to high dim ...., so ?

% since it scales really bad for len, maybe ...
\section{Generalization of The Structure}
\subsection{Satisfing Ring Query}
In this paper, we only discussed the ball query, that is: given a query object $q$ and radius limit $r$, search for set $A$, where $\forall a \in A\; d(a, q)\leq r$. 
% Should I introduce the ring query here or later

However, there are some limitations to this method. First, it cannot do incremental iterative query: if the user is not satisfied with the result for query radius $r$, it is impossible to query for $r_2$ where $r_2>r$ without re-searching the space that is already inspected by the first time query, which created significant unnecessary overhead and yielded poor user experiences. 

The second disadvantage of the ball query is that it might return objects that are not interesting to the user. In the dating app use case, some people might not want to have relationships with others who are too similar to them. In some other use cases of the CMT, such as chemical searching, this could also be an issue. In chemical searching, users might want to get some unexpected molecules instead of a predictable substituted compound. 

A ring query, on the other hand, can solve the above issues. It accepts two radii, $r_{minq}$ and $r_{maxq}$ and returns a set B where $\forall b \in B\; r_{minq}\leq d(b, q)\leq r_{maxq}$. This could satisfy incremental query as 
% we can make a serial of range intervals. (Rephase) 
users can set $r_{minq}$ in the query as the the maxium $r$ in the pervious query. Additionally, a serial of small intervals can be made, and each time user can only query for one of the intervals iteratively. This can decrease the time for each query and increase interactivity and user experience. 

\subsection{Satisfing Set Query}
In this project, we used a rank list to represent each user's interest. However, in some cases, people might put all items that they are not interested in at the end of the list in random order. This might result in the distance between some people being far in the metric space where they actually have similar interests. To solve this issue, a set query might be needed: users select items that they are interested in before they rank them.  However, some set similarity measures are not metrics, and thus the tree needs to be redesigned. Although some set similarity measures are metrics, additional tests might be needed to determine the efficiency of such a search tree. 

% \subsection{Set Query} 
\section{Conclusions}
In this paper, we have tested range query on a search structure, CMT, with \KTDis{}. The results show that the structure can query objects that are within $r$ distance to the query object in practical time with reasonable parameters. Moreover, it scales sublinearly with population size ($N$). Although it scales superlinearly with query distance $r$, if the number of objects found ($nFound$) is controlled, it ``scales'' sublinearly to the $nFound$. Nevertheless, this structure scales poorly to the length of the rank list, so the length of the list should be considered carefully when applied. 

% \subsection{The Effectiveness Of \KTDis{}} 
However, in this paper, we only discussed the scaling of the CMT on \KTDis{}, demonstrating that such an algorithm could efficiently query people that are similar to the target. However, we did not prove that this matching algorithm is effective and the matching can lead to a successful relationship. Future work is needed to establish via psychological or sociological methodologies whether the quality of the matchings produced by the new method does, in fact, exceed that of prior art methods.  
\section*{Acknowledgment}
The computation for this work was performed on the high-performance computing infrastructure provided by Research Computing Support Services and in part by the National Science Foundation under grant number CNS-1429294 at the University of Missouri, Columbia, MO. DOI: https://doi.org/10.32469/10355/69802

ChatGPT, Midjourney, and GitHub Copilot have been lightly used to assist in the preparation of this manuscript. %However, no output of ChatGPT is directly used in this paper. % I still think we might want to acknowledge something like this because it is now in a kind of a gray area 

{}

\end{document}